\documentclass[11pt, conference, letter]{IEEEtran}

\usepackage{graphicx}
\usepackage{amsfonts,mathrsfs,xcolor}
\usepackage{amssymb}
\usepackage{bm}
\usepackage{booktabs}
\usepackage{subcaption}
\usepackage{amsmath}
\usepackage{amsthm}
\usepackage{mathtools}
\usepackage{physics}
\usepackage{lettrine}
\usepackage{hyperref}
\hypersetup{colorlinks=true, urlcolor=blue, linkcolor=blue, citecolor=red}
\usepackage{cleveref}

\graphicspath{{Figures/}}

\newcommand{\R}{\mathbb{R}}
\newcommand{\params}{\bm{\mu}}
\newcommand{\G}{\mathcal{G}}
\newcommand{\N}{\mathcal{N}}
\newcommand{\U}{\mathcal{U}}

\begin{document}

\title{DeepSVM: Learning Stochastic Volatility Models with Physics-Informed Deep Operator Networks}

\author{%
\IEEEauthorblockN{Hakob Chakhoyan}
\IEEEauthorblockA{Department of Mathematics\\Yale University\\ New Haven, CT, USA\\\href{mailto:hakob.chakhoyan@yale.edu}{hakob.chakhoyan@yale.edu}}
\and
\IEEEauthorblockN{Selim Kalici}
\IEEEauthorblockA{Department of Astrophysics\\Yale University\\ New Haven, CT, USA\\ \href{mailto:selim.kalici@yale.edu}{selim.kalici@yale.edu}}
\and 
\IEEEauthorblockN{Kieran A. Malandain}
\IEEEauthorblockA{
Department of Physics\\Yale University\\ 
New Haven, CT, USA\\
\href{mailto:kieran.malandain@yale.edu}{kieran.malandain@yale.edu}} 
}

\maketitle

\begin{abstract}
Real-time calibration of stochastic volatility models (SVMs) is computationally bottlenecked by the need to repeatedly solve coupled partial differential equations (PDEs). In this work, we propose \textit{DeepSVM}, a physics-informed Deep Operator Network (PI-DeepONet) designed to learn the solution operator of the Heston model across its entire parameter space. Unlike standard data-driven deep learning (DL) approaches, DeepSVM requires no labelled training data. Rather, we employ a hard-constrained ansatz that enforces terminal payoffs and static no-arbitrage conditions by design. Furthermore, we use Residual-based Adaptive Refinement (RAR) to stabilize training in difficult regions subject to high gradients. Overall, DeepSVM achieves a final training loss of $\mathbf{10^{-5}}$ and predicts highly accurate option prices across a range of typical market dynamics. While pricing accuracy is high, we find that the model's derivatives (Greeks) exhibit noise in the at-the-money (ATM) regime, highlighting the specific need for higher-order regularization in physics-informed operator learning.
\end{abstract}


\section{Introduction}
\label{sec:intro}
\lettrine{P}{ricing} European options under stochastic volatility models (SVMs) requires solving a two-dimensional parabolic partial differential equation (PDE) for each candidate set of market parameters. This becomes computationally expensive when calibrating the six free parameters of the Heston model \cite{heston1993closed} to real-time, noisy market data, because the PDE solve must be repeated many times \cite{srinivasanFastOptionPricing2025}.

Recent advances in machine learning (ML) have begun to reshape the field of quantitative finance \cite{kellyFinancialMachineLearning, fanDeepLearningSolving2025}. Specifically, Physics-Informed Neural Networks (PINNs) \cite{raissi2019physics} have emerged as a powerful tool for solving financial PDEs without the need for large-scale, labeled datasets \cite{noguerialonsoPhysicsInformedNeuralNetworks2023,hainautOptionPricingHeston2024}. However, standard PINNs typically learn a solution for a fixed set of parameters $\params$ that is determined at model initiation. Our goal is to replace numerical pricing with a neural operator \cite{Lu_2021_DeepONet} that generalizes across parameter regimes, enabling near real-time pricing and calibration.

Crucially, this operator-learning paradigm fundamentally alters the computational economics of model calibration. In a typical volatility calibration routine, an optimizer must solve the pricing PDE thousands of times to fit the model to market quotes. While Finite Difference solvers and standard PINNs incur a high computational cost for every new parameter set queried, DeepSVM amortizes this cost entirely during the training phase. Once trained, the network provides near-instantaneous inference ($\mathcal{O}(1)$ cost) across the global parameter space, effectively removing the bottleneck for real-time risk management and high-frequency trading applications.

The Heston model provides a realistic framework for option pricing by incorporating stochastic volatility, a key feature of financial markets. It assumes that an asset's price $S$ follows a stochastic diffusion, while its variance $\nu$ follows a correlated mean-reverting square-root process:
\begin{align}
\dd{S_t} &= r S_t\dd{t} + \sqrt{\nu_t} S_t\,\dd{W^S_t} \label{eq1:price-sd}\\
\dd{\nu_t} &= \kappa(\theta - \nu_t)\dd{t} + \sigma \sqrt{\nu_t}\dd{W^{\nu}_t},\label{eq1:var-sd}
\end{align}
where the variables are labelled in Table~\ref{tab:param_ranges}, and $W^S, W^\nu$ are correlated Brownian motions with $\langle \dd{W^S}, \dd{W^\nu}\rangle = \rho \dd{t}$. We assume a zero dividend yield ($q=0$) for this study, though the framework generalizes trivially.


In this work, we address the computational challenges of solving the Heston model by introducing \textbf{DeepSVM}. Our three primary contributions are identified as follows:
\begin{enumerate}
    \item We formulate the Heston pricing problem as an \textbf{operator learning} task, allowing for global generalization across market parameters $\params$ rather than a single instance solve.
    \item We introduce a hard-constrained \textbf{ansatz} architecture \cite{lagaris1998artificial} to automatically satisfy the transformed terminal \& boundary conditions, reducing the optimization search space.
    \item We demonstrate that a physics-informed approach, augmented via \textbf{Residual-based Adaptive Refinement (RAR)} \cite{lu2021deepxde}, can achieve high-fidelity pricing (comparable to an FD solver) without the need for expensive ground-truth data generation.
\end{enumerate}
The remainder of this paper is organised thus: \S\ref{sec:problem} details the mathematical formulation; \S\ref{sec:method} describes the DeepSVM architecture; and \S\ref{sec:results} presents numerical validation of our model.

\section{Problem Formulation}
\label{sec:problem}
The arbitrage-free price of a European option, $V(S,\nu,t)$, is governed by the Feynman-Kac (FK) PDE derived from the Heston stochastic processes in Eqs.~\eqref{eq1:price-sd}--\eqref{eq1:var-sd}. In order to facilitate efficient learning via neural networks, which perform better when key regions are centred around the origin \cite{lecun2012efficient}, we transform the coordinates to dimensionless quantities.

\subsection{Log-Moneyness Transformation}
\label{subsec2:log-moneyness}

Standard spot-price coordinates suffer from scaling issues, as the relevant domain for the asset price $S$ shifts with the strike price $K$. Therefore, we introduce the \textit{log-moneyness} coordinate $x$:
\begin{equation}
    x = \ln(\frac{S}{K}),\label{eq2:coord-trans-def}
\end{equation}
along with defining the time-to-maturity $\tau$:
\begin{equation}
    \tau = T-t.\label{eq2:time-to-maturity-def}
\end{equation}
where $T$ is the expiration time. We define the normalized option price $u(x, \nu,\tau) = V(S,\nu,t)/K$. Thus, the Heston PDE transforms into the following parabolic equation for $u$:
\begin{equation}
    \pdv{u}{\tau}-\N[u]=0,\label{eq2:pde-log}
\end{equation}
where the differential operator $\mathcal{N}$ is defined as:
\begin{align*}
    \N[u] &= \left(r - \frac{1}{2}\nu\right)\pdv{u}{x} + \rho\sigma\nu\pdv[2]{u}{x}{\nu} \\
    &+ \frac{1}{2}\sigma^2\nu\pdv[2]{u}{\nu} + \frac{1}{2}\nu\pdv[2]{u}{x} + \kappa(\theta-\nu)\pdv{u}{\nu} - ru.
\end{align*}
This transformation centers the ATM region at $x=0$, creating a scale-invariant domain that simplifies the task for the neural operator we train.

\subsection{Boundary and Terminal Conditions}
\label{subsec2:boundary-terminal-conditions}

The PDE \eqref{eq2:pde-log} must be solved subject to the terminal payoff condition at $\tau=0$ (i.e., $t=T$) and asymptotic boundary behaviours:
\begin{align}
    u(x,\nu,\tau=0) &= \max(e^x-1,0)\equiv\phi(x),\\
    \lim_{x\to\infty}u(x,\nu,\tau) &= e^x-e^{-r\tau},\\
    \lim_{x\to-\infty}u(x,\nu,\tau) &= 0.
\end{align}
Further, as $\nu\to0$, the second-order diffusion terms vanish, and \eqref{eq2:pde-log} simplifies to a hyperbolic transport equation, requiring no boundary condition at $\nu=0$ provided the Feller condition is satisfied.

\subsection{Parameter Space and Domain}
\label{subsec2:param-space-domain}

We formulate the pricing task as an operator learning problem for near-instantaneous pricing  We seek to learn the mapping $\G: \mathcal{P}\to\U$, where $\mathcal{P} \subset \R^5$ is the space of Heston parameters and $\U\equiv\{u: \Omega\to\R\}$ is the solution space over the domain $\Omega = (x,\nu, \tau)$. The specific ranges for the input parameters $\params \in \mathcal{P}$ are detailed in Table~\ref{tab:param_ranges}.

\begin{table}[t]
    \centering
    \caption{Parameter ranges and domain definitions. The \textbf{Model Parameters} form the input to the DeepONet Branch net, while the \textbf{Domain Coordinates} form the input to the Trunk net. The Feller condition $2\kappa\theta > \sigma^2$ is enforced during sampling.}
    \begin{tabular}{llc}
        \toprule
        \textbf{Symbol} & \textbf{Description} & \textbf{Range} \\
        \midrule
        \multicolumn{3}{c}{\textit{Model Parameters} ($\params$)} \\
        $\kappa$ & Mean reversion speed   & $[0.5,\; 3.0]$ \\
        $\theta$ & Long-run variance      & $[0.01,\; 0.20]$ \\
        $\sigma$ & Volatility of variance & $[0.1,\; 1.0]$ \\
        $\rho$   & Spot/variance corr.    & $[-0.95,\; -0.05]$ \\
        $r$      & Risk-free rate         & $[0.00,\; 0.08]$ \\
        \midrule
        \multicolumn{3}{c}{\textit{Domain Coordinates} ($\Omega$)} \\
        $x$      & Log-moneyness $\ln(S/K)$ & $[-2.0,\; 2.0]$ \\
        $\nu$    & Instantaneous Variance   & $[0.01,\; 0.40]$ \\
        $\tau$   & Time-to-maturity         & $[0.0,\;1.0]$\\
        \bottomrule
    \end{tabular}
    \label{tab:param_ranges}
\end{table}

\section{Methodology: {\texttt{DeepSVM}}}
\label{sec:method}
To solve the operator learning problem defined in \S\ref{sec:problem}, we
propose \texttt{DeepSVM}, a physics-informed DeepONet \cite{wang2021learning}
augmented with a hard-constrained ansatz and adaptive sampling.

\subsection{DeepONet architecture}
\label{subsec3:architecture}

\begin{figure*}[ht]
    \centering
    \includegraphics[width=0.75\linewidth]{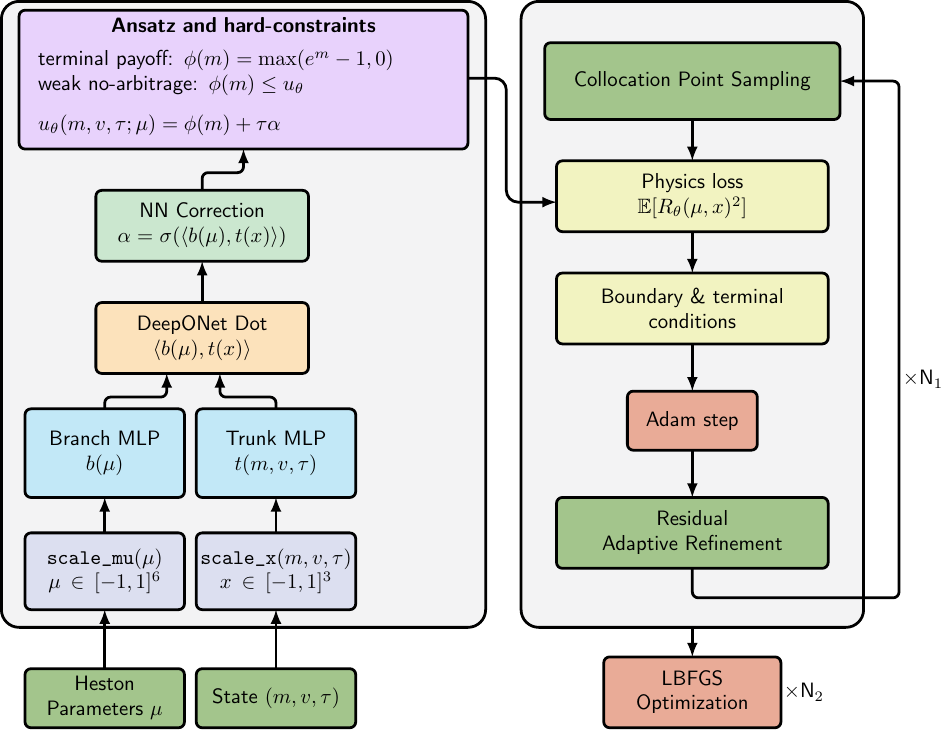} 
    \caption{The \texttt{DeepSVM} architecture. The model combines a DeepONet
    core with a hard-constrained ansatz to enforce the terminal payoff
    condition exactly. Training is stabilized via residual-based adaptive
    refinement (RAR).}
    \label{fig:architecture}
\end{figure*}

We employ a DeepONet \cite{Lu_2021_DeepONet} to approximate the solution
operator \(\mathcal{G}\). As illustrated in Fig.~\ref{fig:architecture},
the architecture consists of two sub-networks:
\begin{enumerate}
    \item \textbf{Branch net:} a multi-layer perceptron (MLP) that encodes
    the input parameters \(\params = (\kappa,\theta,\sigma,\rho,r)\)
    into a latent embedding vector \(\mathbf{b}(\params)\in \mathbb{R}^p\).
    \item \textbf{Trunk net:} an MLP that encodes the spatiotemporal query
    coordinates \((x,\nu,\tau)\) into a basis vector
    \(\mathbf{t}(x,\nu,\tau) \in \mathbb{R}^p\), where
    \(x = \ln(S/K)\) denotes log-moneyness.
\end{enumerate}
The unconstrained output of the network is the inner product of these two
embeddings:
\begin{equation}
    \mathfrak{N}_\theta(\params,x,\nu,\tau)
    = \big\langle \mathbf{b}(\params),
      \mathbf{t}(x,\nu,\tau)\big\rangle,
    \label{eq:network_out}
\end{equation}
where \(\theta\) denotes all trainable parameters of the branch and trunk nets.
Each scalar input component of \((\boldsymbol{\lambda},x,\nu,\tau)\) is
linearly rescaled to lie in the interval \([-1,1]\) before being passed
to the network.
Both branch and trunk MLPs use 4 hidden layers with GELU activations and residual connections between layers, and the correction term $\sigma(\cdot)$ is implemented as a Softplus.

\subsection{Hard-constrained ansatz}
\label{subsec3:ansatz}

In general, vanilla PINNs struggle to exactly satisfy initial or terminal
conditions, leading to error propagation into the interior of the domain
\cite{wang2021learning}. This issue is particularly pronounced in the
Heston stochastic volatility model, where the terminal payoff is not a
stationary solution of the PDE operator. We strictly enforce the terminal
payoff condition at \(\tau=0\) by decomposing the solution into a fixed term
and a learned correction, following the classical ansatz of
\cite{lagaris1998artificial}. Specifically, we define
\begin{equation}
    u_{\text{pred}}(x,\nu,\tau;\params)
    = \phi(x) + \tau\,\sigma\bigl(
      \mathfrak{N}_\theta(\params,x,\nu,\tau)\bigr),
    \label{eq:ansatz}
\end{equation}
where \(\phi(x) = \max(e^x - 1,0)\) is the known terminal payoff defined in
\S\ref{subsec2:boundary-terminal-conditions}, and \(\sigma(\cdot)\) is
a smooth activation function (we use the Softplus nonlinearity).
The multiplicative factor \(\tau\) ensures that as \(\tau \to 0\), the correction term vanishes and \(u_{\text{pred}} \to \phi(x)\) exactly. Furthermore, the use of the positive Softplus function for the correction term enforces the static no-arbitrage condition \(u(x,\nu,\tau) \ge \phi(x)\), ensuring the option price never falls below its intrinsic value.

\subsection{Loss function and training}
\label{subsec3:loss_fn_training}

Let \(\mathcal{N}[u]\) denote the Heston differential operator defined in Eq.~\eqref{eq2:pde-log}. We train the network by minimizing the PDE residual over the domain, with the total loss:
\begin{align}
    \mathcal{L}
    = \mathcal{L}_\textrm{phys}
    + \lambda_b \mathcal{L}_\textrm{bound}
    + \lambda_a \mathcal{L}_\textrm{atm},
    \label{eq:loss_fn}
\end{align}
The PDE residual is a modified penalty formulated to prioritize the highest residuals in the collocation set:
\begin{equation}
\mathcal{L}_\textrm{phys}
= \mathbb{E}[R^2]
 + \lambda_\textrm{max}\mathbb{E}[R^4],
\end{equation}
with $\lambda_\textrm{max}=0.1$ and the squared residual $R^2$ defined as:
\begin{equation}
R^2 = \frac{1}{N}\sum_{i=1}^{N}
\left| \pdv{u_{\textrm{pred},i}}{\tau} - \mathcal{N}[u_{\textrm{pred},i}] \right|^2.
\end{equation}
The term \(\mathcal{L}_{\text{bound}}\) aggregates the boundary-condition errors at the spatial boundaries \(x = x_{\min}\) and \(x = x_{\max}\), while \(\mathcal{L}_{\text{atm}}\) is the same residual loss restricted to a dedicated set of collocation points concentrated in the at-the-money (ATM) region \(x \in [-0.05,0.05]\). We set $\lambda_b = 1.0$ and $\lambda_a = 1.0$ in our experiments.

To combat the high gradients present near the ATM region, we employ \textbf{Residual-based Adaptive Refinement} (RAR) \cite{lu2021deepxde,Wu_2023}. During training, we periodically evaluate the PDE residual on a dense set of candidate points, identify those with the largest residuals, and add them to the active collocation set. This creates a feedback loop in which the model explicitly targets the most difficult regions of the state space.

We generate log-moneyness collocation points by warping a low-discrepancy Sobol sequence \(u_x \in [0,1]\) into the interval \([x_{\min},x_{\max}]\) using a smooth \(\tanh\)-based mapping that concentrates samples near the center of the domain. Define:
\begin{align*}
x_{\mathrm{mid}}  &= \frac{1}{2}\bigl(x_{\min} + x_{\max}\bigr), \\
x_{\mathrm{half}} &= \frac{1}{2}\bigl(x_{\max} - x_{\min}\bigr), \\
z_x               &= 2u_x - 1 \in [-1,1].
\end{align*}
The transformed log-moneyness coordinate is then:
\begin{equation}
    x(u_x) = x_{\mathrm{mid}} 
    + x_{\mathrm{half}} \,\tanh\bigl(\alpha_x z_x\bigr),
\end{equation}
where \(\alpha_x > 0\) controls the degree of clustering around \(x_{\mathrm{mid}}\); we use \(\alpha_x = 2\) in all experiments.

Training proceeds in two stages. First, we run \(10{,}000\) iterations of the Adam optimizer with an initial learning rate of \(10^{-4}\), exponentially decaying the learning rate by a factor of \(1/2\) every \(2{,}000\) steps. We perform a RAR update every \(500\) optimizer steps. The total size of the collocation set is \(200{,}000\) points to account for the high dimensionality of the problem. At each RAR step we draw \(50{,}000\) candidate collocation points, evaluate their residuals, select the \(20{,}000\) points with largest loss, and use them to replace randomly selected points in the active set. In addition, we maintain \(4{,}096\) ATM collocation points and \(2{,}048\) boundary-condition points, which remain fixed throughout the first training stage. In the second stage, we augment the boundary set with an additional \(2{,}048\) points and refine the solution with L-BFGS (memory size \(20\)) for \(5{,}000\) iterations, starting from the Adam solution and reusing the collocation sets from the initial phase.
    
\section{Results \& Discussion}

\label{sec:results}
\begin{figure}
    \centering
    \includegraphics[width=\linewidth]{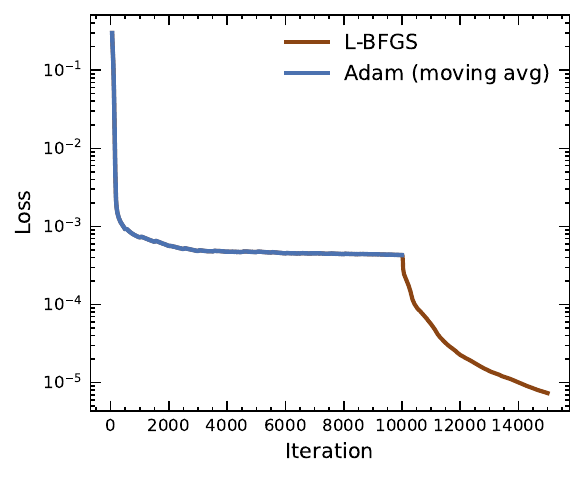}
    \caption{Convergence of the total loss during training. The shaded regions indicate the Adam and L--BFGS phases, respectively.}
    \label{fig:loss}
\end{figure}

\begin{figure*}
    \centering
    \includegraphics[width=\linewidth]{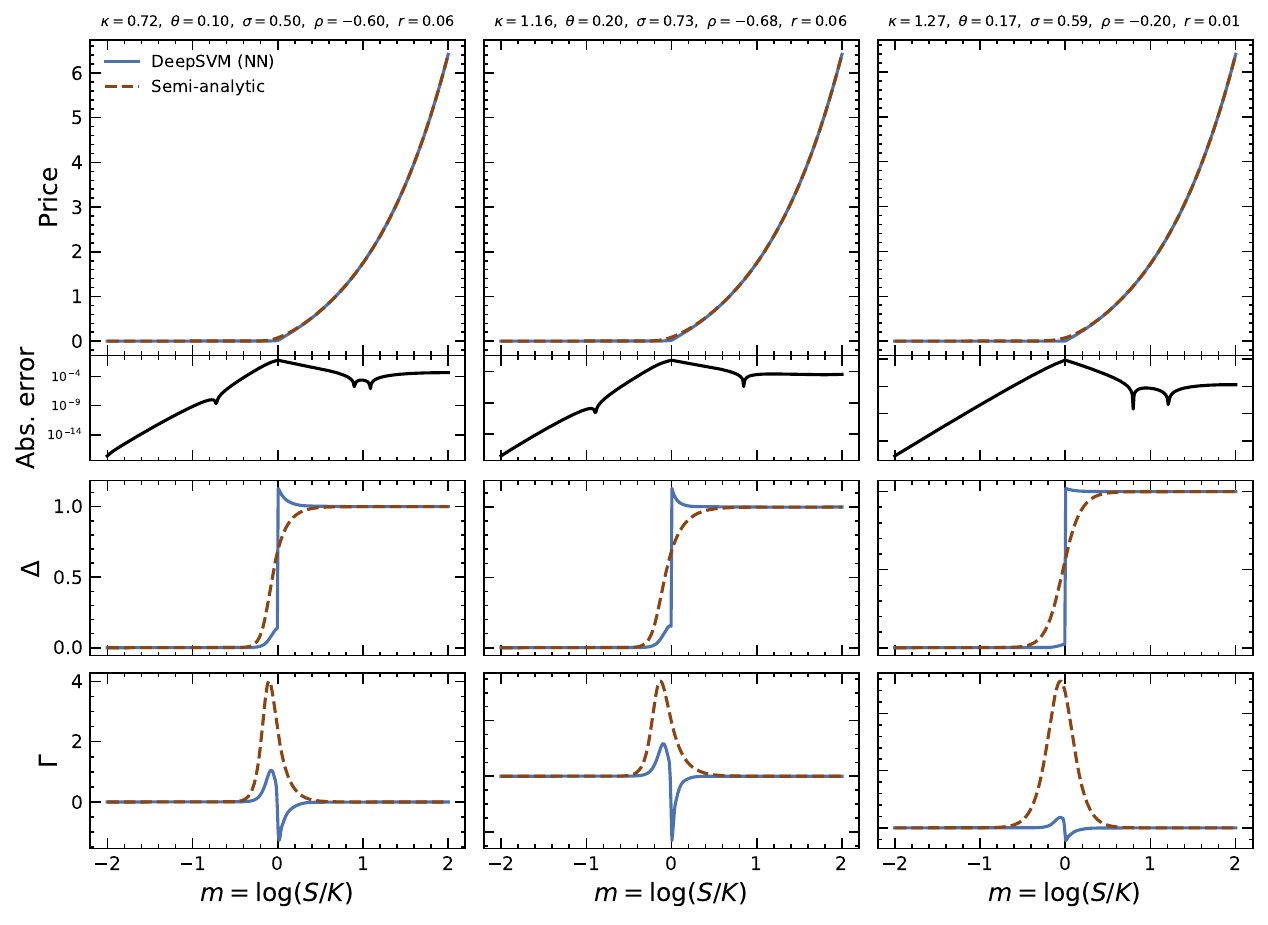}
    \caption{Top two rows: Comparison between DeepSVM and the semi-analytic Heston solution for three randomly selected parameter vectors $\boldsymbol{\mu}$. For each column we show the option price (DeepSVM vs. semi-analytic) and the corresponding absolute pricing error as a function of log-moneyness $x = \ln(S/K)$. Bottom row: For each parameter set, we show the corresponding Greeks (Delta and Gamma) computed via Autodiff from DeepSVM and from the semi-analytic model.}
    \label{fig:semianalytic}
\end{figure*}

\begin{figure*}
    \centering
    \includegraphics[width=\linewidth]{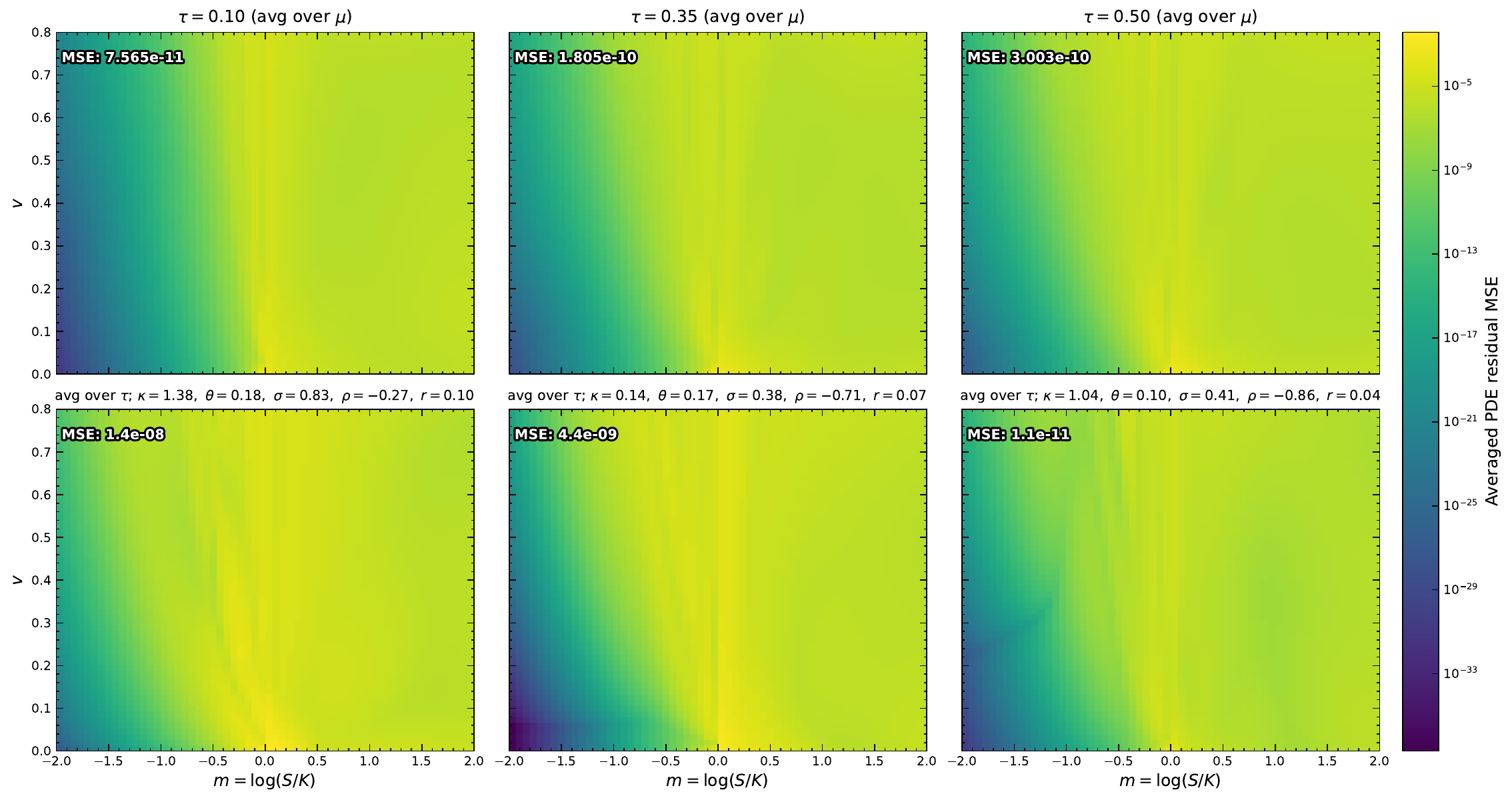}
    \caption{Spatial maps of the PDE residual mean-squared error (MSE) in $(x,\nu)$ space. Top row: Residual MSE averaged over the parameter space $\boldsymbol{\mu}$ for different time-to-maturity slices $\tau$. Bottom row: Residual MSE averaged over $\tau$ for three representative parameter vectors $\boldsymbol{\mu}$. All panels share a common logarithmic color scale.}
    \label{fig:avg_loss}
\end{figure*}

In Figure~\ref{fig:loss} we show the training dynamics of \textbf{DeepSVM}. The Adam phase rapidly decreases the loss before plateauing, where we can observe the impact of the RAR steps every $500$ optimizer steps. Subsequently, the L-BFGS phase refines the solution and reduces the total loss to $\mathcal{O}(10^{-5})$ within $5{,}000$ iterations. Empirically, we found that increasing the number of L-BFGS iterations beyond this point produces only marginal improvements in the objective function, indicating convergence for our current model, sampling scheme, and training routine.

Figure~\ref{fig:semianalytic} assesses the accuracy of DeepSVM against the numerical semi-analytic Heston solution. The top two rows display, for three randomly sampled parameter combinations $\boldsymbol{\mu}$, the predicted call price as a function of log-moneyness $x$ together with the corresponding absolute price error on a logarithmic scale. Across all three chosen parameter sets, DeepSVM reproduces the price surface with high fidelity across the majority of the domain; the largest discrepancies occur in the narrow region near the ATM regime ($x \approx 0$). This confirms that the hard-constrained terminal ansatz employed in DeepSVM, combined with our modified loss and RAR scheme, can drive the pricing error down to a level that is essentially indistinguishable from market noise.

The bottom row of Figure~\ref{fig:semianalytic} highlights a critical observation regarding the derivatives of the learned operator. The Greeks, or sensitivities of the price with respect to state variables, exhibit noise that is not present in the semi-analytic benchmark. The two Greeks used for comparison are Delta ($\Delta$) and Gamma ($\Gamma$). Delta represents the sensitivity of the option price to the underlying asset spot price:
\begin{equation}
    \Delta \equiv \frac{\partial V}{\partial S}
    = e^{-x}\frac{\partial u}{\partial x},
\end{equation}
and Gamma represents the curvature of the option value:
\begin{equation}
    \Gamma \equiv \frac{\partial^2 V}{\partial S^2} = 
    \frac{e^{-2x}}{K}\left[\frac{\partial^2 u}{\partial x^2} - \frac{\partial u}{\partial x}\right].
\end{equation}
While the prices match the semi-analytic benchmark extremely well, the corresponding Greeks computed via automatic differentiation show deviations. Delta exhibits oscillations near the ATM region ($x=0$), deviating from the smooth sigmoid-like profile of the semi-analytic solution. Consequently, Gamma—which involves a second derivative—amplifies these oscillations, leading to locally negative values in regions where convexity should be strictly positive.

Fundamentally, these artifacts arise because the neural network is constrained only by the PDE residual (which couples the value and derivatives) rather than by the derivatives directly. The optimizer finds a solution manifold that minimizes the PDE error but is not necessarily smooth in higher-order derivatives. We experimented with adding a soft constraint penalty to encourage positive values of $\Gamma$ at ATM collocation points, but this did not significantly improve the stability of the learned Greeks and degraded the overall pricing accuracy. This suggests that explicit Sobolev regularization (training on derivative norms) is a necessary direction for future research in financial operator learning.

To understand the structure of the PDE losses, Figure~\ref{fig:avg_loss} shows spatial maps of the residual mean squared error across $(x,\nu)$ space. The top row shows residual maps averaged over uniformly sampled parameter draws $\boldsymbol{\mu}$. We observe that the residual is negligible over the bulk of the domain, with the highest values persisting in localized pockets near the ATM region. The most problematic domain occurs at extremely small variances ATM, which corresponds to the region of maximum non-linearity in the payoff function. The bottom row of Figure~\ref{fig:avg_loss} shows losses averaged over time for a single parameter set for three randomly selected choices of $\boldsymbol{\mu}$. Here we again observe that while the residual is well-behaved globally, ridges of instability persist in the near-ATM regime where the Greeks are also distorted.

The loss analysis demonstrates that DeepSVM functions as an accurate surrogate model for the Heston PDE, achieving highly accurate residuals across large swathes of parameter space. However, standard physics-informed training, even with heavy sampling focused on the ATM regime, does not automatically guarantee smooth higher-order derivatives. This result highlights that while operator learning can solve the pricing problem ($\mathcal{O}(1)$ inference), industrial applications requiring precise hedging parameters (Greeks) will likely require Sobolev-enhanced loss functions.

\section{Conclusion}
\label{sec:conclusion}
For European options in the classical Heston stochastic volatility model, semi-analytic Fourier-based pricing formulas are available and widely used in practice. From a pure pricing standpoint, these methods are typically more efficient than solving the associated pricing PDE.

However, our goal is not to improve on the classical Heston pricer itself, but to develop a general physics-informed operator learning framework (\textbf{DeepSVM}) that can be applied to stochastic volatility models and payoffs without closed-form or semi-analytic solutions. We therefore use the Heston model primarily as a benchmark setting with a trusted reference solution against which we can quantify the accuracy of our learned operator.

In this work, we have:
\begin{enumerate}
    \item Introduced the DeepSVM model, which accurately learns the pricing operator across the entirety of the parameter space corresponding to the Heston stochastic volatility model.
    \item Demonstrated that while pricing accuracy is high, the Greeks computed via automatic differentiation exhibit high-frequency noise when benchmarked against semi-analytic solutions, highlighting a specific regularization challenge in the at-the-money regime.
\end{enumerate}

To address the discrepancies in the Greeks, future work should employ Sobolev training techniques \cite{czarnecki2017sobolevtrainingneuralnetworks}. By differentiating the Heston PDE with respect to the state variables, we can obtain a system of coupled PDEs governing the sensitivities directly. Augmenting the loss function to minimize the residuals of these sensitivity equations would force the network to learn a solution manifold that is smooth in both value and derivative. Practically, these would be equations of the form:
\begin{align*}
\pdv{(\partial_x u)}{\tau} &= \pdv{x}\mathcal{N}[u],\\
\pdv{(\partial_{xx} u)}{\tau} &= \pdv[2]{x}\mathcal{N}[u].
\end{align*}
While this technique would increase the size of the computational graph for each forward pass—since the $\partial_x\mathcal{N}[u]$ term contains third-order mixed derivatives—it represents the most theoretically sound path toward industrial-grade operator learning for finance.

Finally, while this work validates DeepSVM against the semi-analytic Heston benchmark, the true potential of this framework lies in generalized volatility regimes—such as Rough Volatility or high-dimensional Local Stochastic Volatility (LSV) models—where no closed-form solutions exist. In these computationally demanding environments, a pre-trained DeepSVM operator could replace expensive Monte Carlo simulations, offering a viable pathway for real-time exotic option pricing.

\bibliographystyle{IEEEtran}
\bibliography{references}

\end{document}